\documentclass[aps,showpacs,showkeys,supescriptaddress,amsmath,amssymb,12pt]{revtex4}
\usepackage{latexsym}
\usepackage{bm}

\def\s{\sigma}

\usepackage{textcomp}
\usepackage{eurosym}
\usepackage{amsfonts}
\usepackage{array}
\usepackage{amsthm}
\usepackage{bm}
\usepackage[colorinlistoftodos]{todonotes}
\usepackage{mathpazo}
\usepackage{supertabular}
\usepackage{subfig}
\usepackage{amssymb}
\usepackage{eurosym}
\usepackage{amsmath}
\usepackage{epsfig}
\usepackage{graphics}
\usepackage{color}
\usepackage{graphicx}
\def\beq{\begin{eqnarray}}
\def\eeq{\end{eqnarray}}

\usepackage[utf8]{inputenc}
\def\bes{\begin{eqnarray}}
\def\ees{\end{eqnarray}}

\begin{document}

\title{Motion and collision of particles near DST Black holes} 

\author{P. A. Gonz\'{a}lez}
\email{pablo.gonzalez@udp.cl} \affiliation{Facultad de
Ingenier\'{i}a y Ciencias, Universidad Diego Portales, Avenida Ej\'{e}rcito
Libertador 441, Casilla 298-V, Santiago, Chile.}

\author{Marco Olivares}
\email{marco.olivaresr@mail.udp.cl} \affiliation{Facultad de
Ingenier\'{i}a y Ciencias, Universidad Diego Portales, Avenida Ej\'{e}rcito
Libertador 441, Casilla 298-V, Santiago, Chile.}

\author{Ali \"{O}vg\"{u}n}
\email{ali.ovgun@pucv.cl}
\homepage{http://www.aovgun.com}
\affiliation{Instituto de F\'{\i}sica, Pontificia Universidad Cat\'olica de
Valpara\'{\i}so, Casilla 4950, Valpara\'{\i}so, Chile.}

\affiliation{Physics Department, Arts and Sciences Faculty, Eastern Mediterranean University, Famagusta, North Cyprus via Mersin 10, Turkey.}

\author{Joel Saavedra}
\email{joel.saavedra@ucv.cl} 

\affiliation{Instituto de F\'{\i}sica, Pontificia Universidad Cat\'olica de
Valpara\'{\i}so, Casilla 4950, Valpara\'{\i}so, Chile.}

\author{Yerko V\'asquez}
\email{yvasquez@userena.cl}
\affiliation{Departamento de F\'isica y Astronom\'ia, Facultad de Ciencias, Universidad de La Serena,\\
Avenida Cisternas 1200, La Serena, Chile.}

\date{\today }

\begin{abstract}
We consider Deser-Sarioglu-Tekin (DST) black holes as background and we study such the motion of massive particles as the collision of two spinning particles in the vicinity of its horizon. New kinds of orbits are allowed  for small deviations of General Relativity, but  the behavior of the collision is similar to the one observed for General Relativity. Some observables like bending of light and the perihelion precession are analyzed.

\end{abstract}


\maketitle
\flushbottom

\newpage

\tableofcontents

\newpage

\section{Introduction}

The Deser-Sarioglu-Tekin (DST) action is characterized by the action of General Relativity (GR) with additional terms, i.e. non-polynomial terms of the Weyl tensor,  that preserve the (first) derivative order of the GR equations in Schwarzschild gauge and provide non-Ricci flat extensions of GR. DST black holes are obtained using the Weyl technique for pure GR \cite{weyl} which led to rather strange metrics \cite{Deser:2007za}, however GR can be recovered. Their thermodynamics was studied in \cite{Vetsov:2018dte}.\\

The motion of particles in a spherically symmetric spacetime background has been of great interest. It is known that all solar system observations, such as light deflection, the perihelion shift of planets, the gravitational time-delay among other are well described within Einstein’s General Relativity. Also, observational data allow us to fix the parameters of alternative four-dimensional theories, see for instance \cite{Olivares:2013jza, Gonzalez:2015jna}. 
In these regards, we address if  new orbits appear as well as if the DST theory allows us to set the observables better than GR, for small deviation of GR. The study of the motion of  spinning tops (STOPs) in the framework of GR began  with the works of Mathisson \cite{Mathisson} and Papapetrou \cite{Papapetrou}, which was again taken by Tulczyjew \cite{Tulczyjew}, Taub \cite{Taub}, and Dixon \cite{Dixon}. The analysis of spinning particles moving around Schwarzschild black holes was first carried out by Corinaldesi and Papapetrou \cite{Corinaldesi} solving the Mathisson-Papapetrou (MP) equations, and later by Hojman \cite{S. A. Hojman} solving the resulting equations derived from the Lagrangian formalism. An analytic treatment of the trajectories in general Schwarzschild-like spacetimes was carried out in \cite{Zalaquett:2014eia}, 
resulting that a spinning test particle does not follow geodesics, 
due to its interaction 
with tidal forces.\\

On the other hand, it is known that the black holes can act as natural particle accelerators if in a collisional process of two particles near the degenerate horizon of an extreme Kerr black hole, one of the particles has a critical angular momentum by creating a large center of mass (CM) energy \cite{Banados:2009pr}.  
Nowadays, this process is known as the Ba\~nados, Silk and West (BSW) mechanism, which was found for the first time by Piran, Shaham and Katz in 1975 \cite{PS1,Piran:1977dm,PS3}.  The BSW mechanism can be extended to non-extremal black holes \cite{Grib:2010at, Li:2010ej} and to non-rotating charged black holes 
\cite{Zaslavskii:2010aw}. On top of that, it has been argued to be a universal property of rotating black holes \cite{Zaslavskii:2010jd}. The BSW mechanism has been studied for different black hole geometries \cite{BSWBH1, BSWBH2, BSWBH3, BSWBH4}. 
Also, the formation of black holes through the BSW mechanism was investigated in \cite{ata}. It is worth to mention that for the collision of  STOPs in the equatorial plane of a Schwarzschild black hole 
it was found that retrograde trajectories 
can experience significant accelerations, which 
generate 
divergent center-of-mass energies if the STOP collides with another particle moving in the same plane. However, in  order to  reach  such  divergence  the trajectory of the STOP  has  to  pass  from timelike  to  spacelike \cite{Armaza:2015eha}.  In this regards, we address if the DST black hole can acts as a particle accelerators.\\

The manuscript is organized as follows: In Sec. \ref{background} we give a brief review of the DST black hole. Then, we study and discuss about the geodesics in the equatorial plane, and we analyze two observables, in particular the bending of the light and  the perihelion precession. Also, we find the values of the coupling parameter in order to set the observations in both test, in Sec. \ref{GEP}. Then, in Sec. \ref{BSW} we study the collision of spinning particles and we investigate the possibility that the DST black hole acts as a particle accelerator. Finally, our conclusions are in Sec. \ref{conclusion}.

\section{DST black holes}
\label{background}

\begin{figure}[!h]
\begin{center}
\includegraphics[width=80mm]{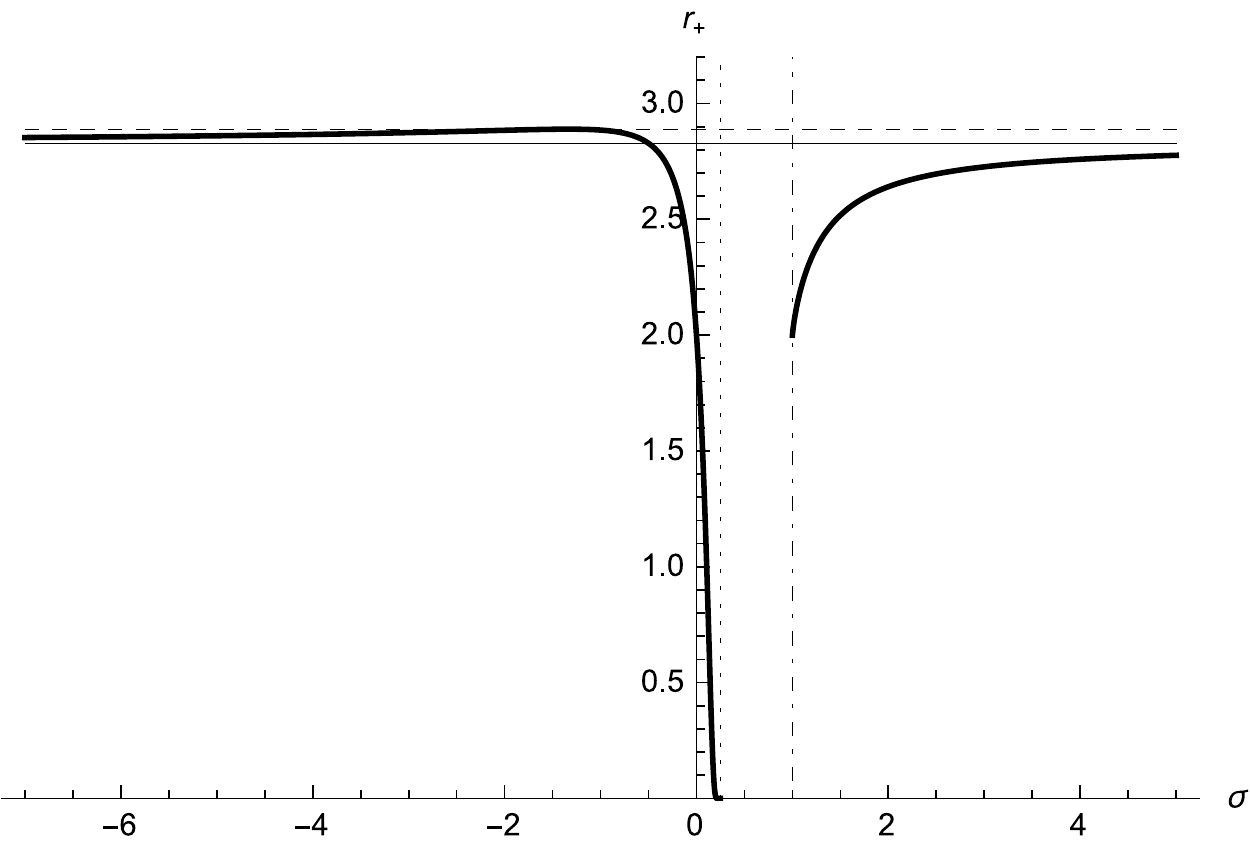}
\end{center}
\caption{The behavior of the DST horizon as a function of $\sigma$ with $r_S=2$.}
\label{HorizonF}
\end{figure}

The action of Deser-Sarioglu-Tekin \cite{Deser:2007za} corresponds to the Einstein action with the addition of non-polynomial terms, which in units of $\kappa = 1$ is given by 
\begin{equation}
I=\frac{1}{2}\int d^4x\sqrt{-g}\left(R+\beta_n \left | tr C^n \right |^{1/n}\right)\,,
\end{equation}
where 
\begin{equation}
tr C^n \equiv C_{ab}{}^{cd}C_{cd}{}^{ef} ... C_{..}{}^{pq} C_{pq}{}^{ab} ... \,,
\end{equation}
being $n$ the number of copies of the Weyl tensor $C$ and $\beta_n$ corresponds to the coupling constant. So, for n=2 and by defining $\sigma=\beta_2/\sqrt{3}$, the above action can be written as 
\begin{equation}
I=\int^\infty_0 dr [(1-\sigma)(arb^{\prime}+b)+3\sigma ab]\,,
\end{equation}
up to boundary terms. Here, primes denote radial derivatives. Thus, for $D=4$, the following metric is solution of the field equations
\begin{equation}
ds^2 = - a(r) \, b^2(r) \, dt^2 + \frac{dr^2}{a(r)} + r^2 \, d\Omega_{2} \;.
\label{met 1}
\end{equation}
where
\begin{equation} 
a(r) = \frac{1 - \s}{1 - 4 \s} + a_1 \, r^{(4 \s - 1)/(1 - \s)} 
\, , \qquad b(r) = b_1 \, r^{3 \s / (\s - 1)} \, , \label{sol} 
\end{equation}
$a_1$ and $b_1$ are integration constants of which $b_1$ is removable by time 
rescaling. Note that for $\s=1$, 
there is no solution at all, for $\sigma=0$ GR is recovered, and for $\sigma=1/4$, $a(r)=\ln(r/r_0)$ and $b(r)=1/r$.  All nonvanishing components of the mixed Weyl tensor are proportional to the single function $X$
\begin{equation}
X(r,t)\equiv \frac{1}{r^2}\left( 2(a-1)-2ra^{\prime}+r^2a^{\prime\prime}\right)+\frac{1}{rb}\left(3ra^{\prime}b^{\prime}-2a(b^{\prime}-rb^{\prime\prime})\right)+\frac{1}{b}\partial_t\left(\frac{1}{a^2b}\partial_t a\right)\,,     
\end{equation}
where primes  denote  radial  derivatives. Also, any scalar of order $n$ in the Weyl tensor $C$ is proportional to $X^n$. Therefore
\begin{equation}
tr C^n = \left(-\frac{1}{3}\right)^n[2+(-2)^{2-n}]X^n\,.
\end{equation}
Note that  the range $1/4 <  \sigma < 1$ is excluded to 
retain the signature. However,  for $a_1< 0$, 
is possible to recover the Schwarzschild metric. In the following we analyze de branch $a_1< 0$, and we consider the functions $a(r)$ and $b(r)$ as
\begin{equation}
a(r)=\frac{1-\sigma }{1-4 \sigma }-\left(\frac{r_{S}}{r }\right)^{\frac{1-4\sigma }{1- \sigma }}\,,
\end{equation}  
\begin{equation}
b(r)=\left(\frac{r}{r_{S} }\right)^{\frac{3\sigma }{\sigma-1}}\,,
\end{equation}  
that are dimensionless, with $r_S$ being the Schwarzschild horizon coordinate. The black hole horizon is given by    
\begin{equation}
r_+(\sigma,r_{S})=r_{S}\left(\frac{ 1-4 \sigma}{1-\sigma }\right)^{\frac{1-\sigma }{1-4 \sigma }}\,,
\end{equation}
that depends on the parameter $\sigma$ and the Schwarzschild's horizon. The event horizon, as a function of $\sigma$, has a maximum value $r_+^{(max)}\approx 1.44r_S$ when
\begin{equation}
\sigma_1= -\frac{e - 1}{4 - e}\approx -1.34\,.
\end{equation}
Also, when $\sigma \rightarrow \pm\infty$, $r_+ \rightarrow \sqrt{2}r_S$,
see Fig. \ref{HorizonF}. 

\section{Geodesics in the equatorial plane}
\label{GEP}
The Lagrangian associated to the metric (\ref{met 1}) is
\begin{equation}\label{tl4}
  2\mathcal{L}=- a\,b^2\dot{t}^2+
  \frac{\dot{r}^2}{a}+r^2\dot{\theta}^2
  +r^2\sin^2\theta\dot{\phi}^2\,,
\end{equation}
where $\dot{q}=dq/d\lambda$, and $\lambda$ is an affine parameter along the geodesic that
we choose as the proper time $\tau$ for particles. Since the Lagrangian (\ref{tl4}) is
independent of the  coordinates ($t, \phi$), then their
conjugate momenta ($\Pi_t, \Pi_{\phi}$) are conserved. The equations of motion can be obtained from
\begin{equation}
 \dot{\Pi}_{q} - \frac{\partial \mathcal{L}}{\partial q} = 0\,,
\label{w.10} 
\end{equation}
which yield
\begin{equation}
\dot{\Pi}_{t} =0 , \quad \dot{\Pi}_{r} =
-{\dot{t}^{2}\over 2}{d(a\,b^2)\over dr}+ 
{\dot{r}^{2}\over 2}{d(a^{-1})\over dr}
+r\,\dot{\theta}^2
+r\sin^2\theta\dot{\phi}^2\,,
\label{w.11a}
\end{equation}
\begin{equation}
\dot{\Pi}_{\theta} = r^2\sin\theta \cos\theta\, \dot\phi^2, \quad
\textrm{and}\quad \dot{\Pi}_{\phi}=0\,.
\label{w.11b}
\end{equation}
where $\Pi_{q} = \partial \mathcal{L}/\partial \dot{q}$
are the conjugate momenta to the coordinate $q$, in particular 
\begin{equation}
\Pi_{t} = -a\,b^2\,\dot{t} , \quad \Pi_{r}= {\dot{r}\over a}\,,
\label{w.11c}
\end{equation}
\begin{equation}
\Pi_{\theta} = \dot{r}^{2}\,\dot{\theta}\,, 
\quad 
\textrm{and}\quad \Pi_{\phi}
= r^2\sin^2\theta\,\dot{\phi}\,.
\label{w.11d}
\end{equation}
So, by considering the motion of neutral particles on the
equatorial plane: $\theta=\pi/2$ and $\dot\theta =0$, we obtain 
 \begin{equation}
\Pi_{t} = -a\,b^2\, \dot{t}\equiv -E\,, \quad \Pi_{r}= {\dot{r}\over a}\,,
\quad \Pi_{\phi}=r^2\,\dot{\phi} \equiv L\,,
\label{w.11c}
\end{equation}
where $E$  and $L_{\phi}$ are integration constants dimensionless. 
Now, by using equations (8) and (11), the Lagrangian can be rewritten in the following form:
\begin{equation}
2\mathcal{L}\equiv -m=-\frac{E^2}{a\,b^2}+ \frac{\dot{r}^2}{a}+
  {L^2\over r^2}\,.
\end{equation}
So, by normalization, we shall consider that $m = 1$ for massive particles and $m = 0$ for photons.
We solve the above equation for $\dot{r}^2$ in order to obtain the radial equation, which allows us to
characterize the possible movements of test particles without an explicit solution of the equation
of motion in the invariant plane, and we obtain
\begin{eqnarray}
&&\left(\frac{dr}{d\lambda}\right)^{2}={1\over b^2} \left[ E^2-V(r)\right]\,, \\
\label{w.12}
&&\left(\frac{dr}{d t}\right)^{2}= {a^2b^2\over E^2}\left[E^2-V(r)\right]\,,\\
\label{w.13}
&&\left(\frac{dr}{d\phi}\right)^{2}=  {r^{\,4}\over L^2\,b^2}\left[E^2-V(r)\right]\,,
\label{w.14}
\end{eqnarray}
where $V(r)$ is the effective potential
given by
\begin{equation}\label{tl8}
  V(r)=a\,b^2\left(m+\frac{L^2}{r^2}\right)\,.
\end{equation}

\subsection{Null geodesics}

\subsubsection{Radial motion}

\begin{figure}[!h]
\begin{center}
\includegraphics[width=90mm]{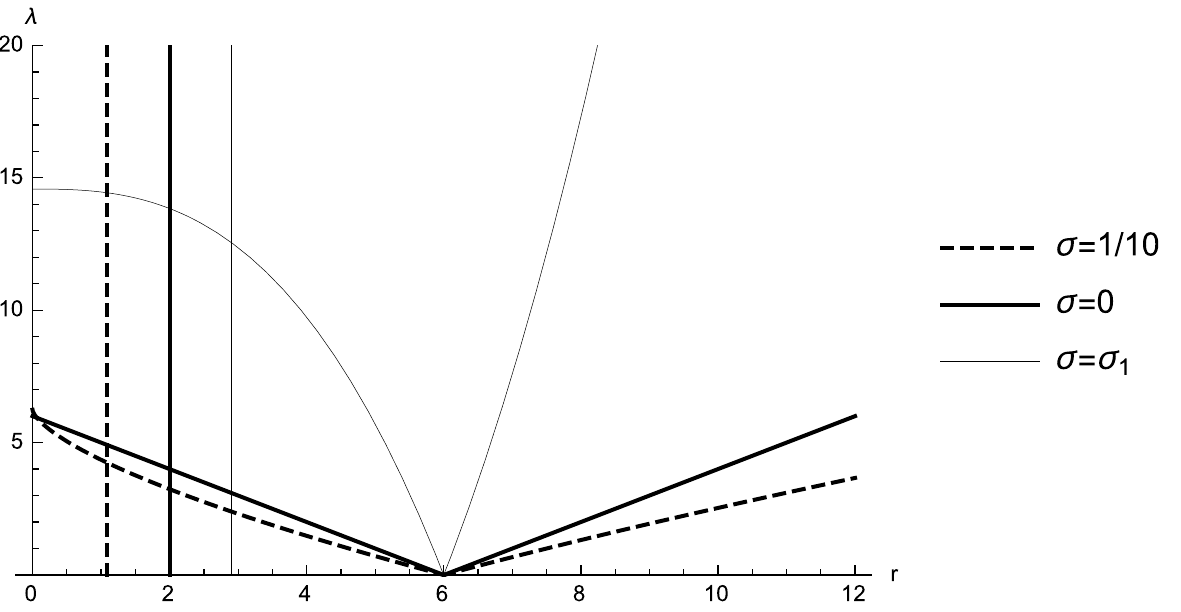}
\end{center}
\caption{The behavior of the affine parameter $\lambda$ as a function of $r$, for different values of $\sigma$, with $E=1$, $\rho_0=6$, $r_{S}=2$. Vertical lines correspond to the event horizon for different $\sigma$ values. }
\label{Tau}
\end{figure}

The radial motion corresponds to a trajectory with null angular
momentum (or zero impact parameter). In this case, the photons are destined 
to escape at the infinity or fall into the black hole due to  the effective potential is $V(r)=0$. Also, Eqs. (\ref{w.12}) and (\ref{w.13}) reduce to 
\begin{equation}
\pm\frac{dr}{d\lambda}= {E\over b},
\label{mr.1}
\end{equation}
and
\begin{equation}
\pm \frac{dr}{dt}=a\,b\,,
\label{mr.2}
\end{equation}
respectively. The sign $+$ ($-$) corresponds to photons that escape (falling) from the event horizon. Now, choosing the initial conditions for the photons as $r=\rho_0$
when $t=\lambda=0$, the Eq. (\ref{mr.1}) yields
\begin{equation}
\lambda(r)=\pm\frac{r_S}{E}\left( {1-\sigma\over 1-4\sigma}\right)\left[ \left (\frac{r}{r_s}\right )^ {1-4\sigma\over 1-\sigma}-\left(\frac{\rho_0}{r_S}\right)^ {1-4\sigma\over 1-\sigma}\right] \,.
\label{mr.3}
\end{equation}
Thus, the photons
arrive to the event horizon in a finite $\lambda$ parameter, which can be observed in Fig. \ref{Tau}. Notice that for photons plunge to the horizon, with the same energy. The photon in the neighborhood of a DST black hole with positive $\sigma$ ($0<\sigma<1/4$) reaches a point (outside the horizon) in a less affine parameter that a photon in the neighborhood of a DST black hole with a negative $\sigma$. 

\begin{figure}[!h]
\begin{center}
\includegraphics[width=90mm]{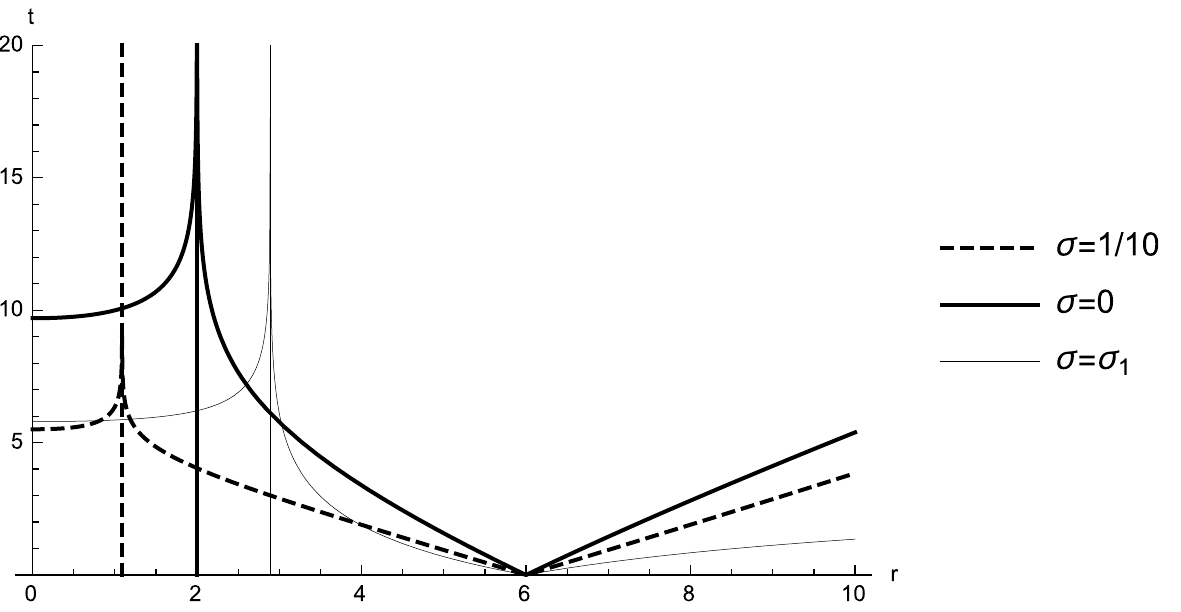}
\end{center}
\caption{The behavior of the coordinate time $t$ as a function of $r$, for different values of $\sigma$, with $\rho_0=6$, and $r_S=2$. Vertical lines correspond to the event horizon for different  $\sigma$ values.}
\label{Tiempo}
\end{figure}

On the other hand, performing the change of variables $x=(r/r_+)^{\nu +1}$, and  integrating Eq. (\ref{mr.2}) leads to
\begin{equation}
t(r)= \pm r_s  \left(\frac{r_+}{r_S}\right)^{1-\nu}\left(B[x;z+1,0]-B[x_0;z+1,0]\right)\,,
\label{mr.5}
\end{equation}
where $B[x;\bar{\alpha},\bar{\beta}]$ corresponds to the Beta function, $z=(1-\nu)/(1+\nu)$ and $(1-\sigma)/(1-4\sigma)=1/(1+\nu)$. Notice that the solution for the coordinate time does not depend on the energy of the photon. In Fig. \ref{Tiempo}, we can observe that the photons in the coordinate time not cross the horizon of the DST black holes.

\newpage

\subsubsection{Angular motion}

\begin{figure}[!h]
\begin{center}
\includegraphics[width=90mm]{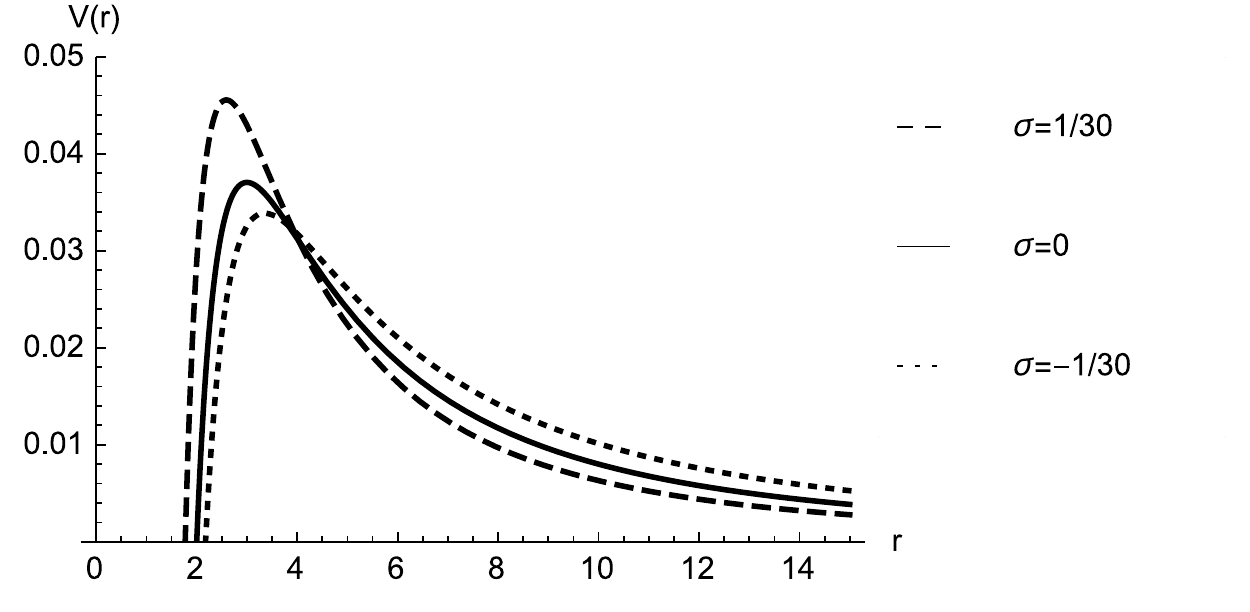}\\
\includegraphics[width=50mm]{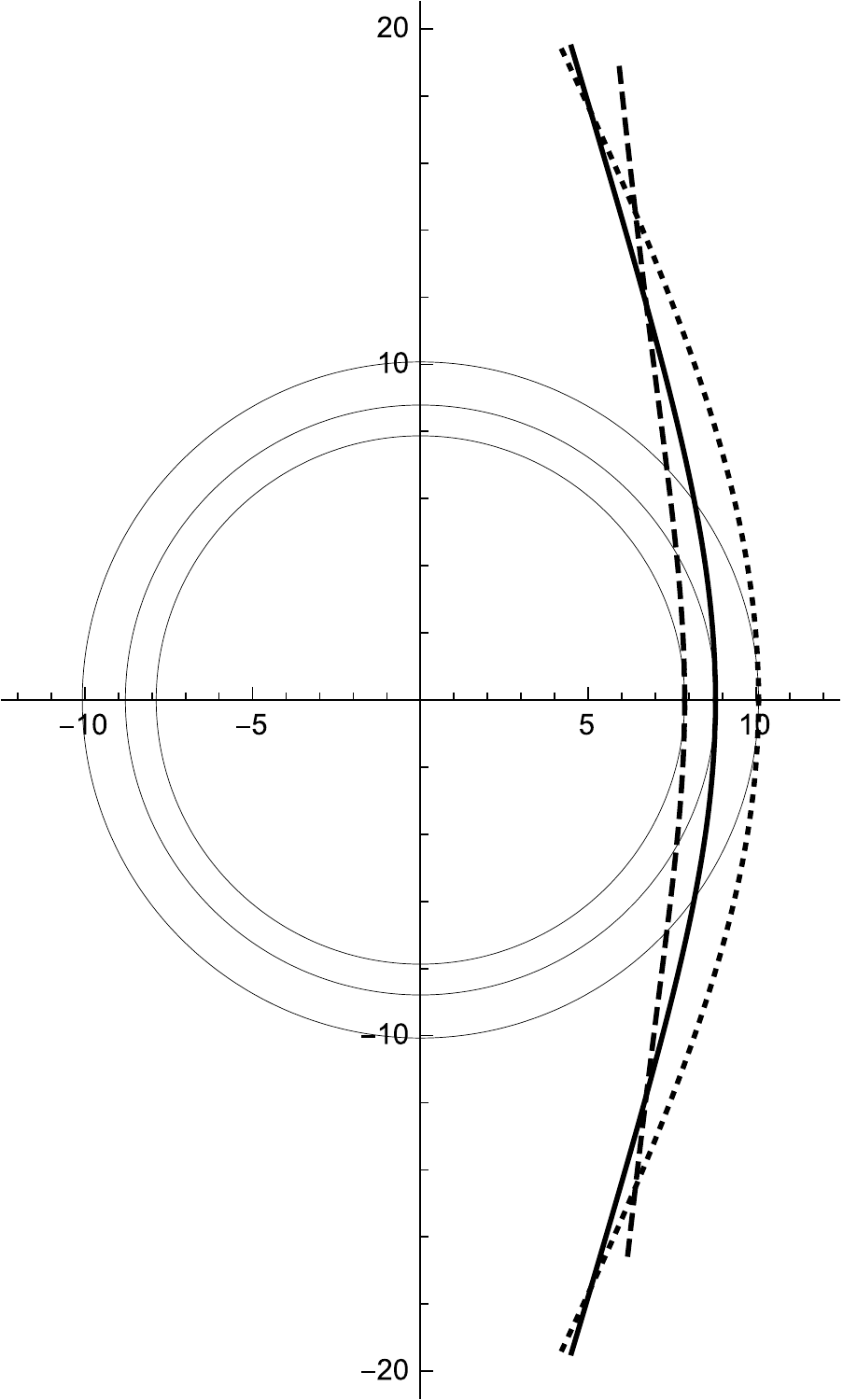}
\end{center}
\caption{ The top panel shows the behavior of the effective potential for photons V(r) as a function of $r$, for different values of $\sigma$, with $r_S=2$ and $L=1$. The bottom panel shows the trajectories for photons with $E^2=0.01$, and different values of $\sigma$.}
\label{PFoton}
\end{figure}

The effective potential for photons and their trajectories, are plotted in Fig. \ref{PFoton}.  The effective potential presents a maximum value, that correspond to a unstable circular orbit with a radius
given by 
\begin{equation}
r_{U}= r_S\left(\frac{3(1-4\sigma)}{2(1-\sigma) (1+2\sigma)}\right)^{\frac{1-\sigma}{1-4
\sigma}}\,.
\end{equation}
The classical result for the Schwarzschild's spacetime ($r_U=3M$) is obtained when $\sigma=0$, and $r_S=2M$. It is worth mentioning that there is a maximum in the potential only for $-1/2<\sigma<1/4$. On the other hand, the potential shows a different behavior for $\sigma \leq -1/2$. \\

Now, based on the impact parameter values $\beta\equiv L/E$, we give a brief qualitative description of the allowed angular motions for photons plotted in Fig. \ref{PFoton}.  First, we can observe a \emph{capture zone}, if $0<\beta<\beta_{c}$, where $\beta_{c}=L/E_c$ with $E_c=V(r_U)$, where the photons fall on the horizon, depending on initial conditions, and its cross section, $\bar{\sigma}$, in these geometry is \cite{wald}
  \begin{equation}\label{mr51}
    \bar{\sigma}=\pi\,\beta_c^2\,.
  \end{equation}
 Also, we can observe a \emph{critical trajectories}, if $\beta=\beta_{c}$, where the photons can stay in one of the unstable circular orbit of radius  $r_{U}$.
  Therefore, the photons that arrive from an initial distance
  $r_i$ ($r_+ < r_i< r_U$)
  can asymptotically fall to a circle of radius $r_{U}$.
  The proper period in such orbit is
  \begin{equation}\label{p1}
  T_{\tau}=\frac{2\pi r_U^2}{L}\,.
  \end{equation}
  It results to be the same as the one in the Schwarzschild case \cite{shutz}, when $\sigma=0$.  Also, the coordinate period is given by
  \begin{equation}\label{p2}
  T_t=2\pi\beta_c\,.
  \end{equation}
Finally, there is a \emph{deflection zone}, if $\beta_{c} <\beta_d<\infty $, where the photons come from  infinity  to a distance $r=r_{d}$ (which is solution of the equation $V(r_d)=E$), then return to the infinity. \\

\begin{figure}[!h]
\begin{center}
\includegraphics[width=90mm]{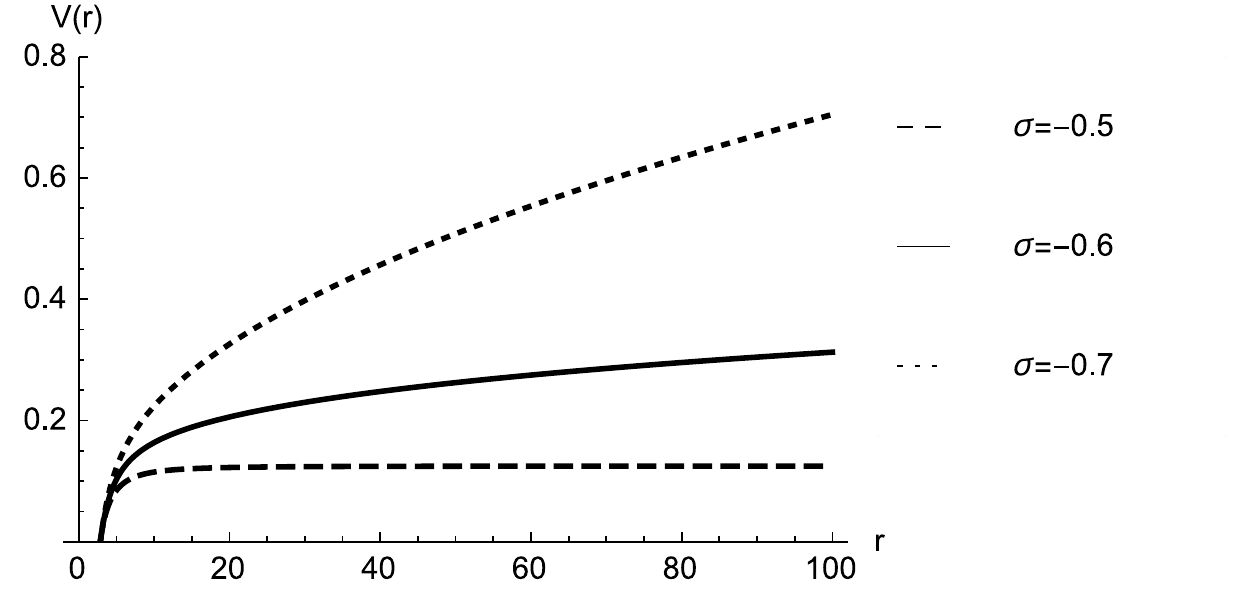}
\end{center}
\caption{The behavior of the effective potential for photons V(r) as a function of $r$, for $\sigma \leq -1/2$, with $r_S=2$ and $L=1$.}
\label{PFoton2}
\end{figure}

As we mentioned, the potential shows a different behavior for $\sigma \leq -1/2$. While that for $\sigma=-1/2$, the potential tends to $L^2/2$, for $\sigma < -1/2$ the potential shows that all trajectories allowed have a return point and then plunge into the black hole, see Fig. \ref{PFoton2}.

\subsubsection{Bending of light}
In this section we will follow the procedure establish in Ref. \cite{Straumann}. So, Eq. (\ref{w.14}) for photons is
\begin{equation}
\left(\frac{dr}{d\phi}\right)^2=\frac{r_S^{2\nu}}{\beta^2} r^{4-2\nu}-\mu r^2+r_S ^{1/\mu}r^{1-\nu}\,,
\end{equation}
where $\beta$ is the impact parameter, $\nu=3\sigma/(\sigma-1)$ and $\mu=(1-\sigma)/(1-4\sigma)$. By using the change of variables $r=1/u$, the above equation can be written as
\begin{equation}
\label{u}
\left(\frac{du}{d\phi}\right)^2=\frac{r_S^{2\nu}}{\beta^2} u^{2\nu}-\mu u^2+r_S ^{1/\mu} u^{3+\nu}\,.
\end{equation}
Notice that for $\sigma=0$, the above equation is reduced to the classical equation of Schwarzschild for the motion of photons for $r_S=2M$
\begin{equation}
\left(\frac{du}{d\phi}\right)^2=\frac{1}{\beta^2}-u^2+2M u^{3}\,.
\end{equation}
So, the derivate of Eq. (\ref{u}) with respect $\phi$ yields 
\begin{equation}
u^{\prime\prime}+\mu u =\frac{r_S^{1/\mu}}{2}(3+\nu)u^{2+\nu}+\frac{ \nu r_S^{2\nu} }{\beta^2}u^{2\nu-1}\,,
\end{equation}
where ${}^\prime$ denotes the derivative with respect to $\phi$. Now, neglecting the last term, we obtain
\begin{equation}
u=\frac{1}{\beta}\sin(\sqrt{\mu}\phi)+\frac{\epsilon}{\beta^{2+\nu}2^{1+\nu/2}\mu}\left(1 +\frac{2+\nu}{6}\cos(2\sqrt{\mu}\phi)\right)\,, 
\end{equation}
where $\epsilon = r_S^{1/\mu}(3+\nu)/2$. For large $r$ (small $u$), $\phi$ is small, and we may take $\sin(\sqrt{\mu}\phi)\approx \sqrt{\mu}\phi$ and $\cos(2\sqrt{\mu}\phi)\approx 1$. In the limit $u \rightarrow 0$, $\phi$ approaches $\phi_{\infty}$, with
\begin{equation}
\phi_{\infty}=-\frac{(3+\nu)(8+\nu)r_S^{1/\mu}}{6\mu^{3/2}\beta^{1+\nu}2^{2+\nu/2}}\,.
\end{equation}
Therefore, for the DST black holes the deflection of light $\hat{\alpha}$ is equal to $2\left |\phi_{\infty}\right |$ and yields 
\begin{equation}\label{GB1}
\hat{\alpha}=\frac{(3+\nu)(8+\nu)}{3\mu^{3/2}2^{2+\nu/2}}\left(\frac{r_s}{\beta} \right)^{1+\nu}\,.
\end{equation} 
Notice that for $\sigma=0$, and $r_S=2M$ we recovered the classical result of GR, that is,  $\hat{\alpha}=4M/\beta$. The first observational value of deflection light was measured by Eddington and Dyson in the solar eclipse of March 29, 1919. For Sobral expedition this value is
$\hat{\alpha}_{Obs.}= 1.98 \pm 0.16''$ and $\hat{\alpha}_{Obs.}= 1.61 \pm 0.40''$ for the Principe expedition \cite{Straumann}. Nowadays, the parameterized post-Newtonian (PPN) formalism introduce the phenomenological parameter  $\gamma$, that characterizes the contribution of space curvature to gravitational deflection,  in this formalism  the deflection angle $\hat{\alpha}=0.5(1+\gamma)1.7426$, and currently $\gamma= 0.9998 \pm 0.0004$ \cite{Shapiro:2004zz}. So, $\hat{\alpha}=1.74277''$ for $\gamma = 0.9998+0.0004$ and $\hat{\alpha}=1.74208''$ for $\gamma=0.9998-0.0004$. \\

It is worth to mention that there is a discrepancy between the theoretical value predicted by GR and the observational value. So, by attributing this discrepancy to small deviations of Schwarzschild's spacetime, we can attribute such discrepancy to $\sigma$. Therefore, $-8.97241\times10^{-6}<\sigma<3.41708\times10^{-6}$, in order to mach with the observational results.

\subsection{Time like geodesics }
In this section, we will study the motion of massive particles and the perihelion precession. In the following, we fix $m=1$.

\subsubsection{Radial geodesics}

\begin{figure}[!h]
\begin{center}
\includegraphics[width=90mm]{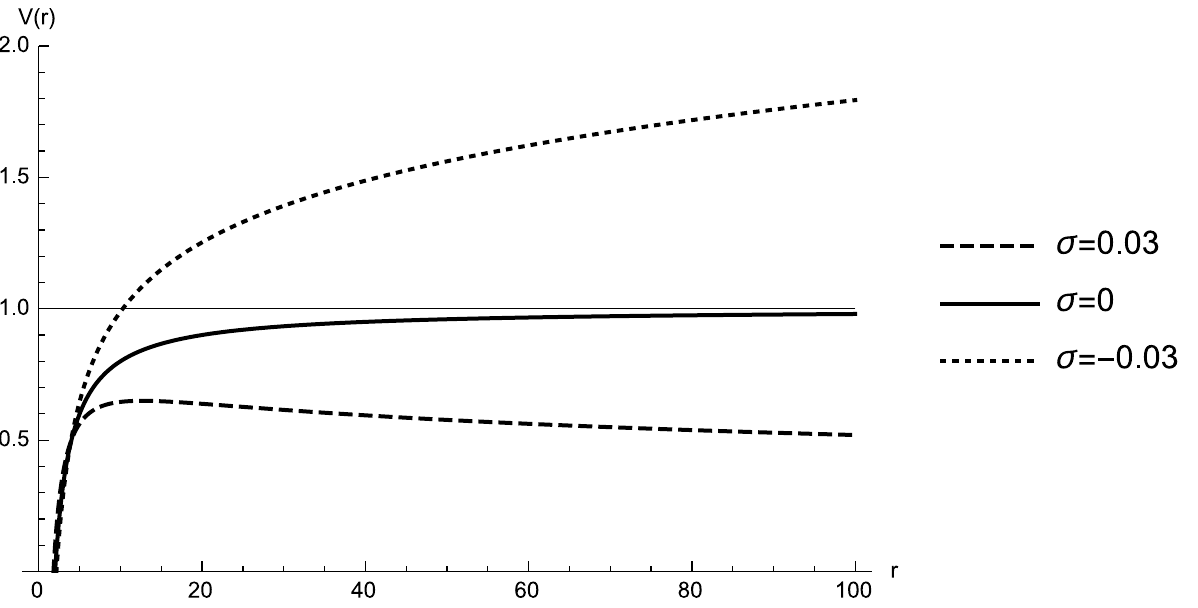}
\end{center}
\caption{The behavior of the effective potential for particles V(r) as a function of $r$, for different values of $\sigma$, $r_S=2$ and $L=0$. For $\sigma=0$ the potential $V(\infty)=1$, for $\sigma<0$ the potential $V(\infty)=\infty$, and for $0<\sigma<1/4$ the potential $V(\infty)=0$.  }
\label{PPPR}
\end{figure}

The effective potential for particles ($L=0$) is plotted in Fig. \ref{PPPR}. Notice that, for $\sigma =0$,  there are two kind of trajectories. One of them is the bounded  trajectory ($E<1$), which has a return point and plunge into the horizon. The other one, is the unbounded trajectory ($E\geq 1$), which can escape at the infinity or plunge into the black hole. For $\sigma <0$, we observe that the allowed trajectories are bounded. Interestingly,  for $0<\sigma<1/4$, the potential has a maximum value $V(r_u)\equiv E_u^2$, at the unstable equilibrium point ($r_u$), that it is not present in GR ($\sigma = 0$), and it can be obtained from the derivative of Eq. (\ref{tl8}) with respect to $r$. This unstable equilibrium point,  is given by
\begin{equation}
r_{u}= r_S\left(\frac{(1-4\sigma)(1+2\sigma)}{6\sigma(1-\sigma) }\right)^{\frac{1-\sigma}{1-4
\sigma}}\,.
\end{equation}
Also, for this range of values of $\sigma$, there are three kind of trajectories. The first, are the critical trajectories, of first and second kind, which are allows for particles with $E=E_u$. The trajectories of first kind are characterized for particles that incoming from the infinity to the unstable equilibrium point, asymptotically. The trajectories of second kind are characterized for particles that incoming from a distance, $r<r_u$, to the unstable equilibrium point, asymptotically. For, particles with $E>E_u$, the trajectories are unbounded, and for $E<E_u$, we observe that is allow a frontal scattering, that is characterized for particles that incoming from infinity to a radial distance of closest approach, and come back to the infinity. The frontal scattering 
for charged particles was studied in Ref. \cite{Villanueva:2015kua}.     

 \newpage

\subsubsection{Angular geodesics}

\begin{figure}[!h]
\begin{center}
\includegraphics[width=90mm]{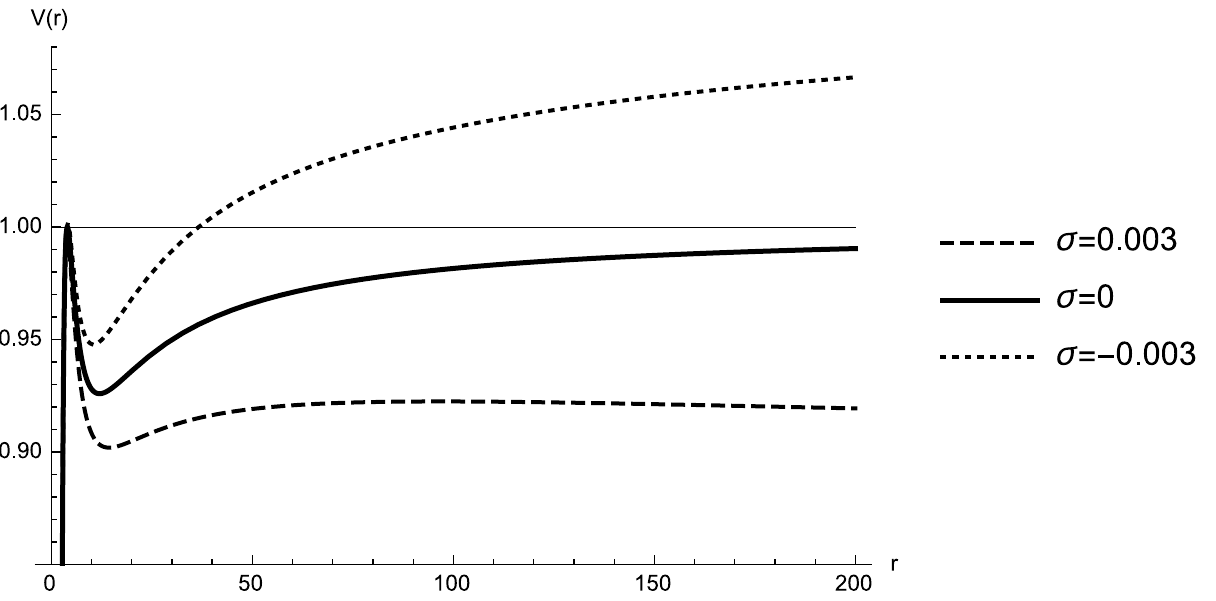}
\end{center}
\caption{The behavior of the effective potential for particles V(r) as a function of $r$, for different values of $\sigma$, with $r_S=2$ and $L=4$. For $\sigma=0$ the potential $V(\infty)=1$, for $\sigma<0$ the potential $V(\infty)=\infty$, and for $0<\sigma<1/4$ the potential $V(\infty)=0$.}
\label{PP}
\end{figure}

\begin{figure}[!h]
\begin{center}
\includegraphics[width=80mm]{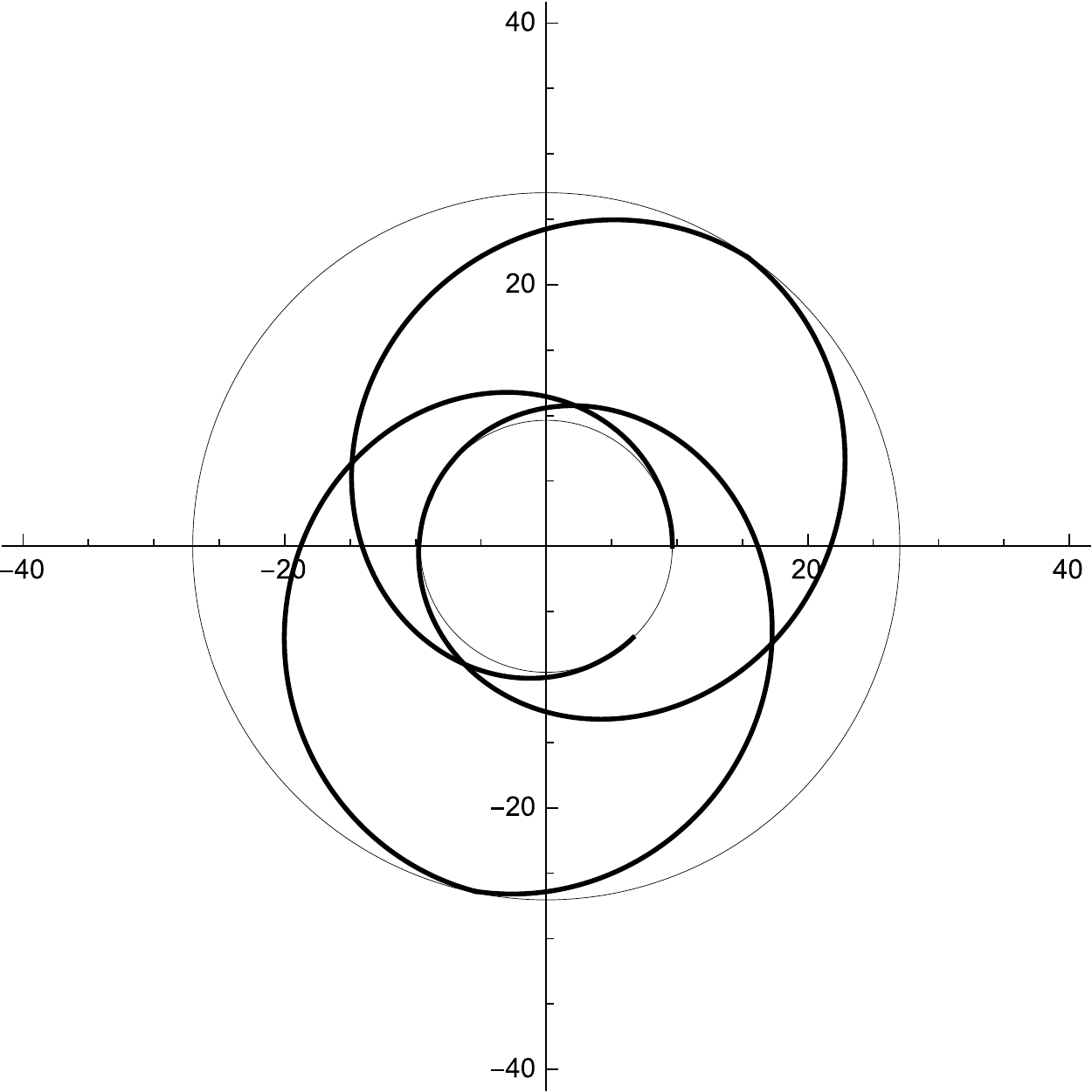}
\includegraphics[width=80mm]{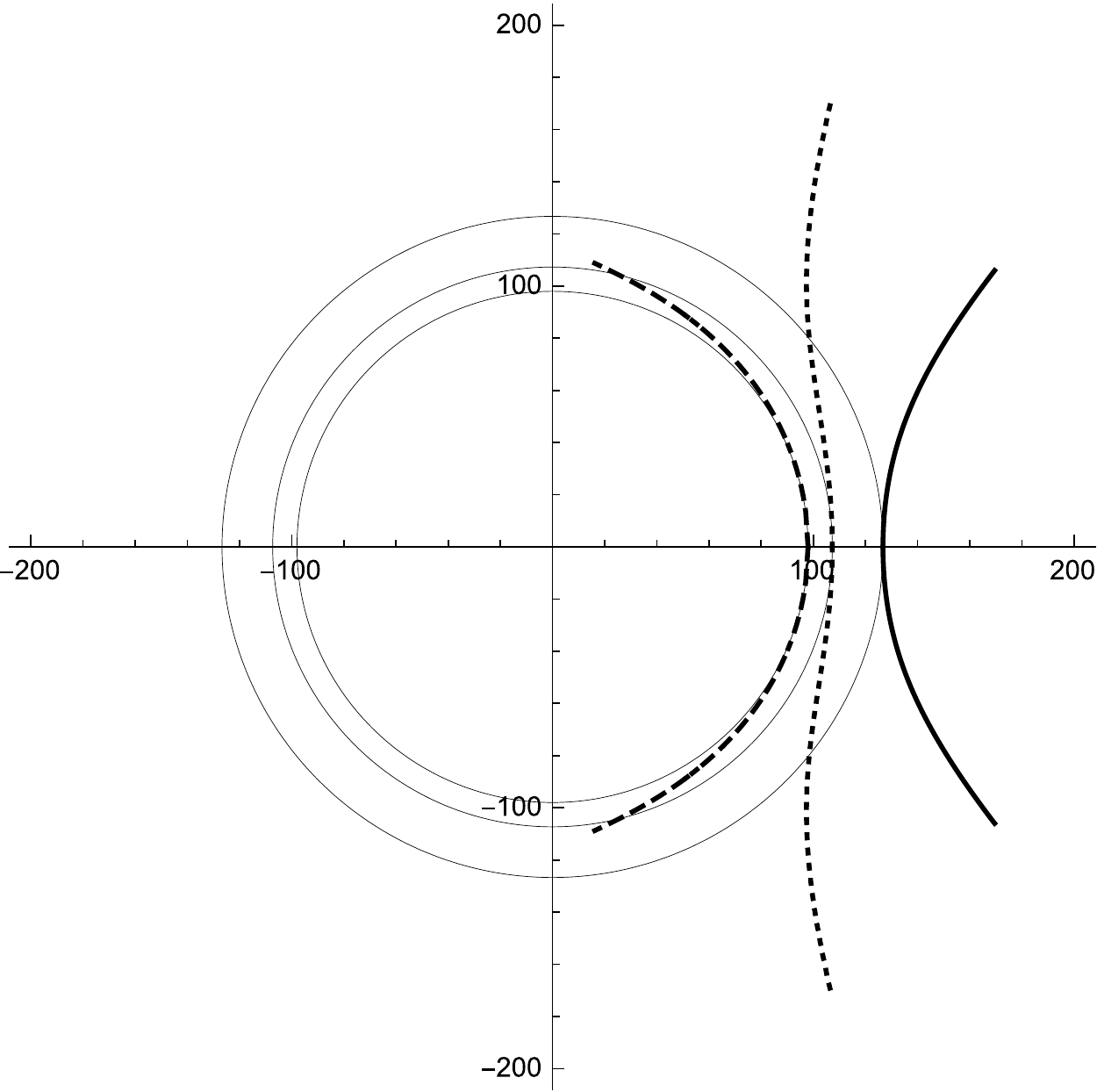}
\end{center}
\caption{Trajectories of particles with angular momentum $L=4$  in the background of DST black hole with $r_S=2$ and $\sigma=0.003$. Left panel for bounded orbits like planetary orbit with energy $E^2=0.91$. Right panel  for the scattering of neutral particles with different values of energy, the continuos line corresponds to $E^2 =0.922$ (repulsive scattering), the dashed line corresponds to $E^2=0.92247$ (atractive scattering) and the dotted line corresponds to $E^2=0.9224$.}
\label{PP12}
\end{figure}

The effective potential for particles with positive angular momentum is plotted in Fig. \ref{PP}. Note that, for 
the cases that have been analyzed the effective potential has a maximum and a minimum value, which corresponds to a unstable and stable circular orbit, respectively. Also, there are bound orbits like planetary orbits, as in GR \cite{chandra}, for instance see left panel of Fig. \ref{PP12} for a positive $\sigma$. Moreover, for $\sigma<0$, all the trajectories are bounded due to $V(\infty)=\infty$. It is worth mention that these kind of orbits have the same behavior that the time like orbits for Schwarzschild AdS black hole \cite{Cruz:2004ts}. Interestingly, for $0<\sigma<1/4$, the spacetime allow two unstable circular orbits and one stable circular orbit. In the right panel of Fig. \ref{PP}, we show the scattering of neutral particles with $E<1$, that are not present in  the Schwarzschild spacetime, which can be a repulsive scattering  or an attractive scattering.

\newpage

\subsubsection{Perihelion precession}
The previous analysis of effective potential for particles showed that there are planetary orbits, which  allow us study the perihelion precession. So, we follow the treatment performed by Cornbleet \cite{Cornbleet}, which
allows us to derive the formula for the advance of the
perihelia of planetary orbits. The starting point is to
consider the line element in unperturbed Lorentz coordinates
\begin{equation}
ds^2 = -dt^2 + dr^2 + r^2(d\theta^2 + sin^2 \theta d\phi^2)~,
\end{equation}
together with line element (\ref{met 1}). So, considering
only the radial and time coordinates in the binomial
approximation, and $b(r)\approx 1$, when $\sigma \rightarrow 0$ in the Newtonian limit. So, the transformation gives
\begin{equation}
d\tilde{t}\approx \sqrt{\mu}\left(1-\frac{1}{2\mu} \left(\frac{r_S}{r}\right)^{\frac{1}{\mu}}\right) dt~, 
\end{equation}
\begin{equation}\label{r}
d\tilde{r}\approx \frac{1}{\sqrt{\mu}}\left(1+\frac{1}{2\mu} \left(\frac{r_S}{r}\right)^{\frac{1}{\mu}}\right) dr~. 
\end{equation}
We will consider two elliptical orbits, one the classical
Kepler orbit in $(r, t)$ space and a DST black hole orbit in
$(\tilde{r},\tilde{t})$ space. Then, in the Lorentz space $dA = \int_0^{\mathcal{R}}rdrd\phi=\mathcal{R}^2d\phi/2$, and hence
\begin{equation}
\frac{dA}{dt}=\frac{1}{2}\mathcal{R}^2\frac{d\phi}{dt}~,
\end{equation}
which corresponds to Kepler's second law. For the
DST black hole case we have
\begin{equation}\label{A}
d\tilde{A}=\int_0^{\mathcal{R}}rd\tilde{r}d\phi~,
\end{equation}
where $d\tilde{r}$ is given by Eq. (\ref{r}). So, we can write (\ref{A}) as
\begin{eqnarray}\label{A2}
\nonumber d\tilde{A}&=&\frac{1}{\sqrt{\mu}} \int_0^{\mathcal{R}}r \left(1+\frac{1}{2\mu} \left(\frac{r_S}{r}\right)^{1+\nu}\right)drd\phi\\
&&= \frac{1}{2\sqrt{\mu}}\left( {\mathcal{R}}^2+\frac{r_S ^{1+\nu}{\mathcal{R}}^{1-\nu}}{\mu(1-\nu)}\right)d\phi~.
\end{eqnarray}
Therefore, applying the binomial approximation 
we obtain
\begin{eqnarray}
\nonumber \frac{ d\tilde{A}}{d\tilde{t}}&=&\frac{1}{2\sqrt{\mu}}\left( {\mathcal{R}}^2+\frac{r_S ^{1+\nu} {\mathcal{R}}^{1-\nu}}{\mu(1-\nu)}\right)\frac{d\phi}{d\tilde{t}}\\
&&\approx \frac{\mathcal{R}^2}{2\mu}\left( 1+\frac{(3-\nu)}{2\mu(1-\nu)}\left(\frac{r_S}{{\mathcal{R}}}\right)^{1+\nu}\right) \frac{d\phi}{dt}~.
\end{eqnarray}
So, using this increase to improve the elemental angle
from $d\phi$ to $d\tilde{\phi}$. Then, for a single orbit
\begin{equation}\label{so}
\int_0^{\Delta\tilde{\phi}}d\tilde{\phi}=\int_0^{\Delta\phi=2\pi}\frac{1}{\mu}\left( 1+\frac{(3-\nu)}{2\mu(1-\nu)}\left(\frac{r_S}{{\mathcal{R}}}\right)^{1+\nu}\right)  d\phi\,.
\end{equation}
Now, as the polar form of an ellipse is given by
\begin{equation}\label{ell}
\mathcal{R}=\frac{l}{1+\epsilon cos\phi}\,,
\end{equation}
where $\epsilon$ is the eccentricity and $l$ is the semi-latus rectum; and by plugging Eq. (\ref{ell}) into Eq. (\ref{so}), we obtain
\begin{eqnarray}
\Delta\tilde{\phi}=\frac{1}{\mu}\left( 2\pi+\frac{(3-\nu)r_S^{1+\nu}}{2\mu(1-\nu)}\int_0^{2\pi} \left( \frac{1+\epsilon cos\phi }{l}\right)^{(1+\nu)} d\phi\right)~,
\end{eqnarray}
which at first order yields 
\begin{eqnarray}
\Delta\tilde{\phi}\approx \frac{1}{\mu}\left( 2\pi+\frac{\pi(3-\nu)}{\mu(1-\nu)}\left(\frac{r_S}{l}\right)^{1+\nu}\right)~.
\end{eqnarray}
Notice that $\sigma=0$, and $r_S=2M$ we recover the classical result of GR.  It is worth to mention that there is  a discrepancy between
the observational value of the precession of perihelion for Mercury,
$\Delta\tilde{\phi}_{Obs.}=5599.74\,(arcsec/Julian-century)$ and the total
$\Delta\tilde{\phi}=\Delta \phi_{eq}+\Delta \phi_{pl}+\Delta \phi_{obl} = 5603.24\,(arcsec/Julian-century)$, where the term $\Delta \phi_{eq}$ is caused by the general precession in longitude, the term $\Delta \phi_{pl}$ is caused by the gravitational tugs of the other planets, and  the term $\Delta \phi_{obl}$ is caused by the oblateness of the Sun \cite{Olivares:2013jza}.
Which, is possible attribute to a DST theory with $\sigma = 1.244 * 10 ^{-9}$. 

\section{Collisions of spinning particles near DST black holes}
\label{BSW}

\begin{figure}[!h]
\begin{center}
\includegraphics[width=70mm]{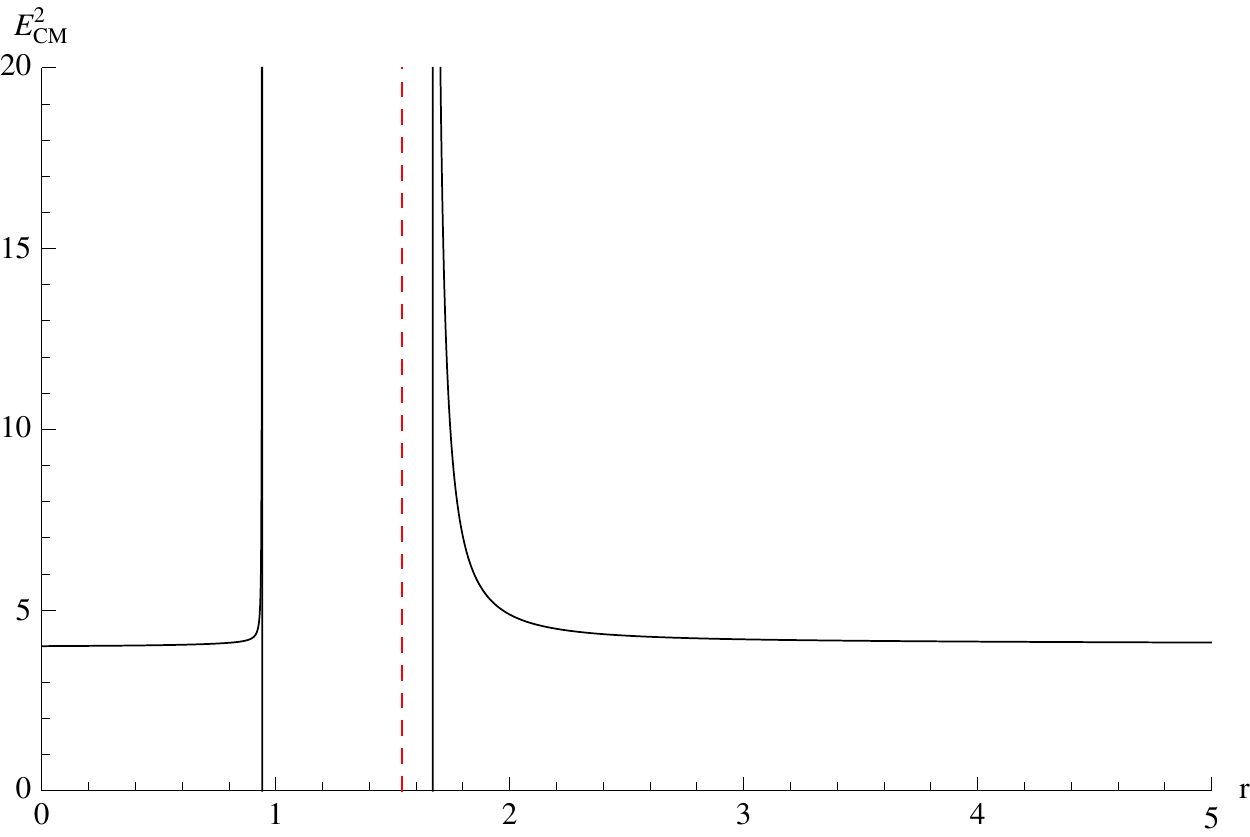}
\end{center}
\caption{The behavior of the $E_{CM}^2$ as a function of $r$, for $a_1=-2$, $b_1=1$, $s_1=1$, $s_2=2.5$, $e_1=e_2=j_1=j_2=1$, $m_1=m_2=1$ and $\sigma=0.1$. The red line corresponds to the event horizon radius $r_H \approx 1.54$. $E_{CM}^2$ diverges at the spin-related radius}
\label{CM}
\end{figure}

\begin{figure}[!h]
\begin{center}
\includegraphics[width=50mm]{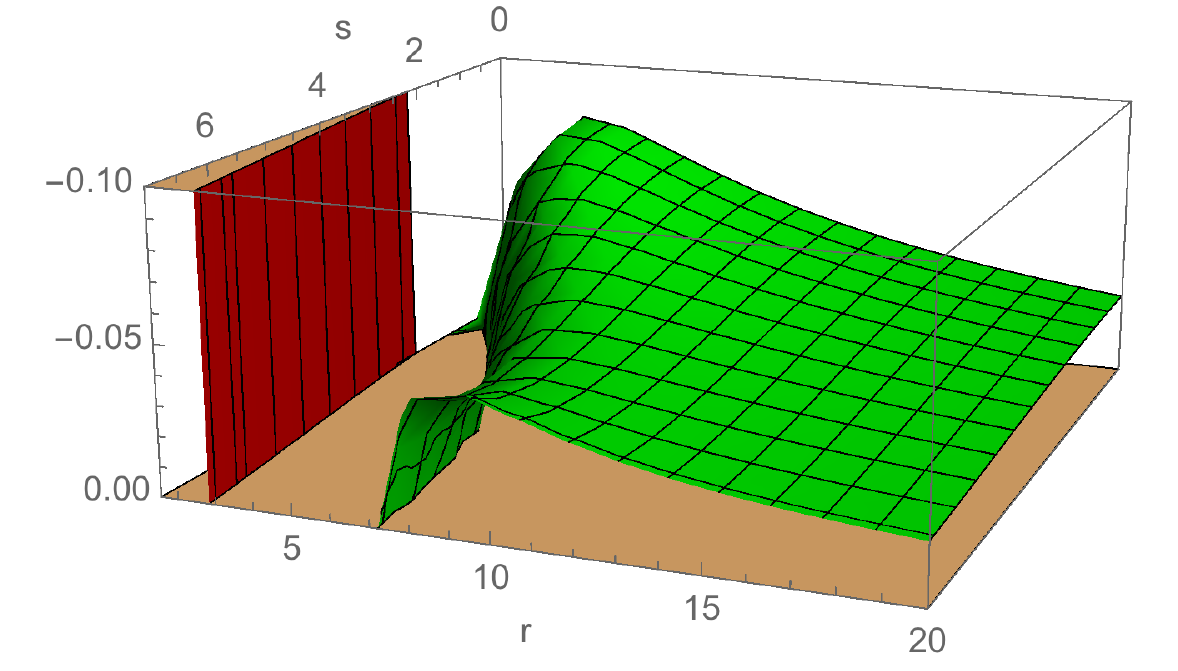}
\includegraphics[width=50mm]{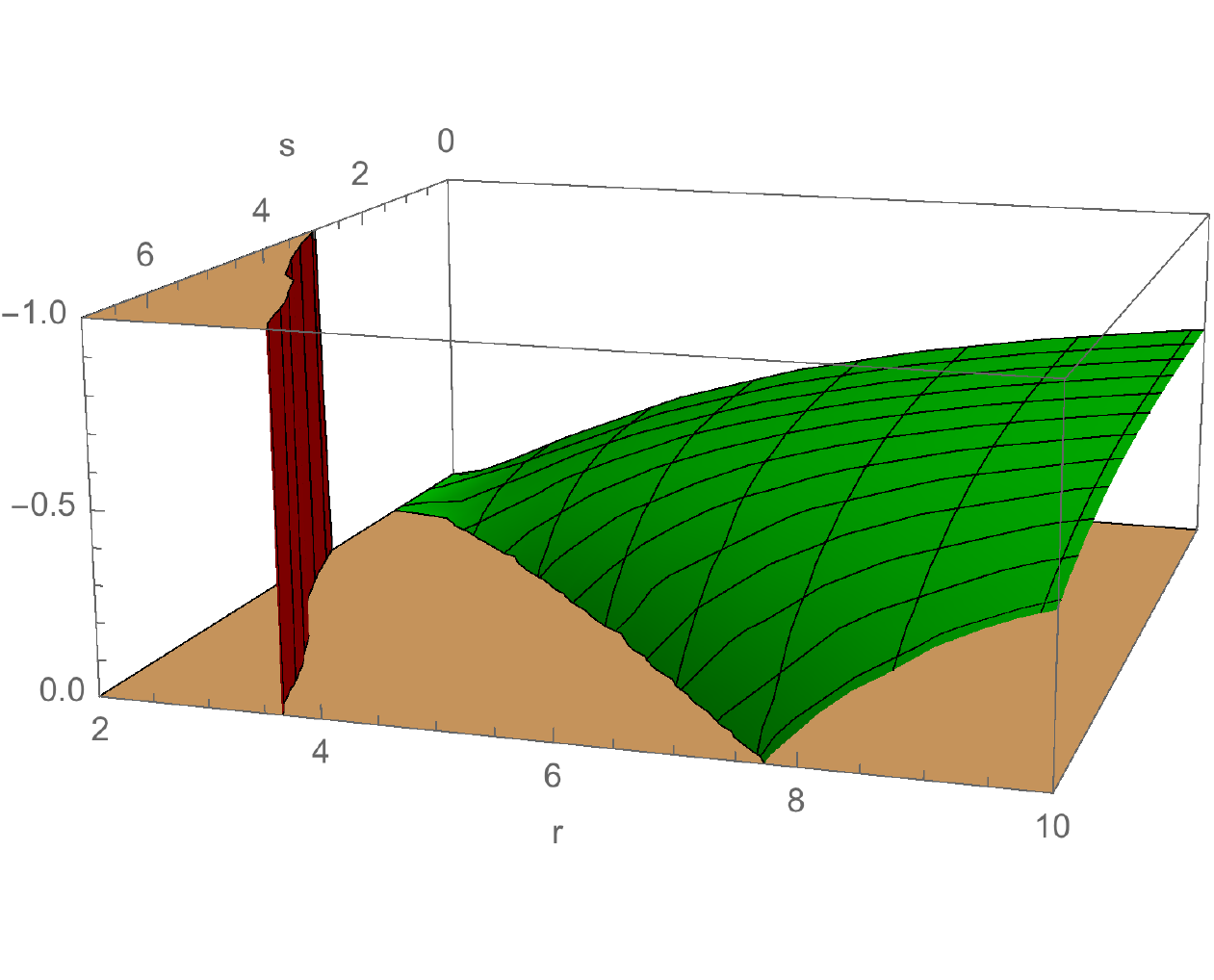}
\includegraphics[width=50mm]{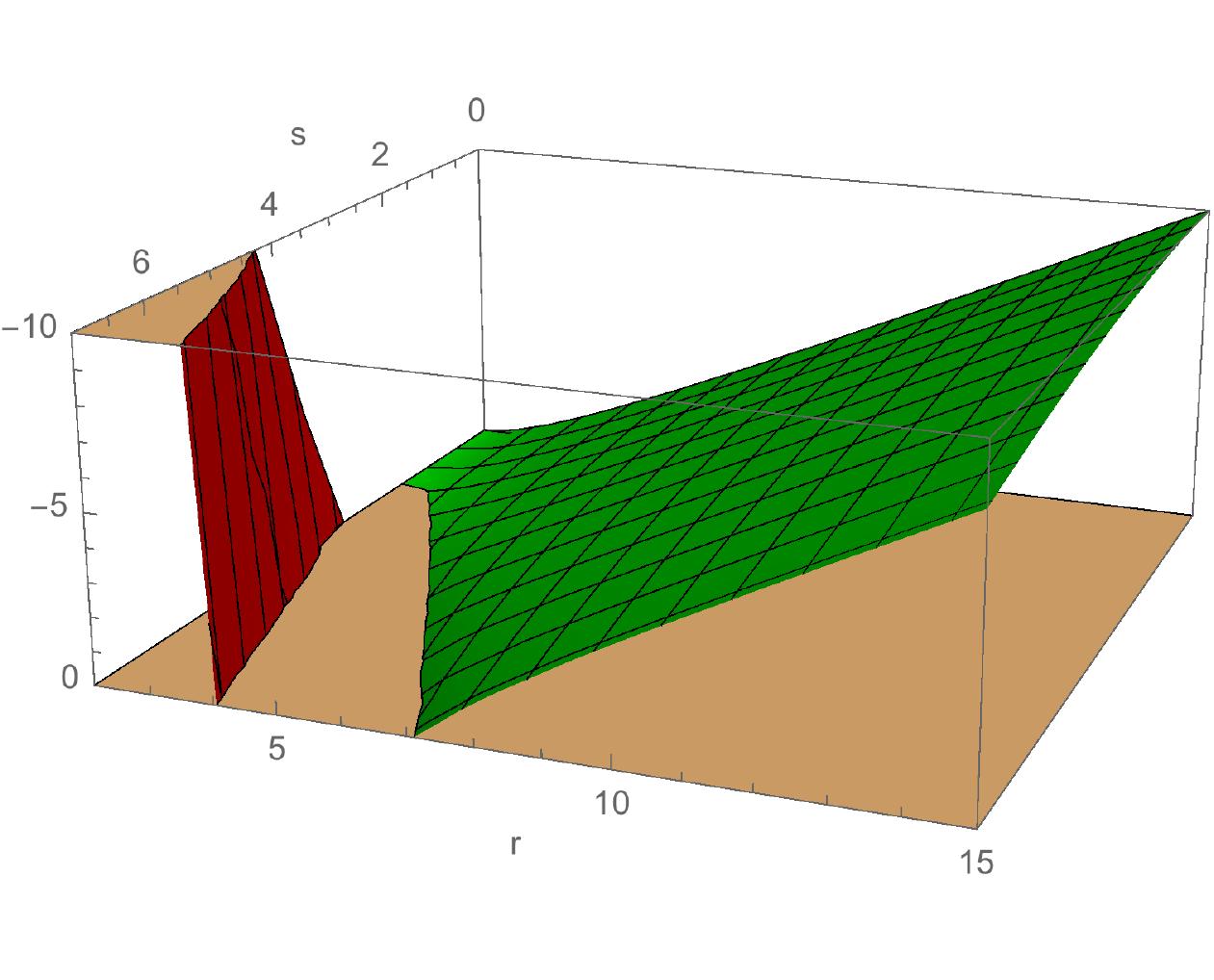}
\end{center}
\caption{The behavior of the velocity square $u_{\mu}u^{\mu}/ \left( u^t \right)^2$ and $\Sigma$ as a function of $r$ and $s$, with $a1=-2M$, $M=1$, $b_1=1$, $E=1$, $j=-0.5$; left panel for $\sigma=0.1$, central panel for $\sigma=0$, and right panel for $\sigma=-0.1$.}
\label{uu}
\end{figure}

The equations of motion derived from the Lagrangian theory for a spinning particle is given by \cite{S. A. Hojman, R. Hojman}
\begin{equation} \label{motion}
\frac{DP^\mu}{D\tau}=-\frac{1}{2}R^\mu_{\nu\alpha\beta}u^\nu S^{\alpha\beta}\,,
\end{equation}
\begin{equation} \label{motion1}
\frac{DS^{\mu\nu}}{D\tau}=S^{\mu\beta}\sigma_\beta\\^\nu-\sigma^{\mu\beta}S_\beta\\^\nu=P^\mu u^\nu-u^\mu P^{\nu}\,,
\end{equation}
where $D/D\tau \equiv u^\mu \nabla_\mu$ is the 
covariant derivative along the velocity vector $u^\mu$, $\tau$ is an affine parameter, $P^\mu$ is
the canonical momentum, $R^\mu_{\nu\alpha\beta}$ is the Riemann tensor, $u^\mu=dX^\mu/d\tau$  is the tangent vector to the trajectory, $S^{\mu\nu}$ is the canonical spin tensor, and $\sigma^{\mu\nu}$ is the angular velocity. Spinning test particles in cosmological and static spherically symmetric spacetimes have been studied in \cite{Zalaquett:2014eia}. Also, the collision of spinning particles near a Schwarzschild black hole was analyzed in \cite{Armaza:2015eha}. In the following, we will consider the motion of spinning particles in the equatorial plane, that is, $\theta=\pi/2$ and $P^\theta=0$.
The modulus of the antisymmetric spin tensor and the mass of the particle are conserved quantities and are given respectively by
\begin{equation}
S^2=\frac{1}{2}S_{\mu \nu}S^{\mu \nu}\,,
\end{equation}
\begin{equation}
m^2=-P_{\mu}P^{\mu}\,.
\end{equation}
Other constants of motion are given by
\begin{equation}
Q_{\xi}=P^{\mu} \xi_{\mu}-\frac{1}{2}S^{\mu \nu} \nabla_{\nu}\xi_{\mu}\,,
\end{equation}
where $\xi^{\mu}$ is a Killing vector of the space-time. The conserved quantity associated to the Killing vector $\partial_t$ corresponds to the energy of the top and is found to be
\begin{equation}
E=a(r)b^2(r)P^t-\frac{1}{2}S^{tr} \left( a'(r)b^2(r)+2 a(r) b(r) b'(r) \right)\,.
\end{equation}
While, the conserved quantity associated to $\partial_\phi$ corresponds to the angular momentum of the top
\begin{equation}
J=r^2 P^{\phi}+r S^{r \phi}\,.
\end{equation}
Also, the Tulczyjew constraint restrict the spin tensor to generate rotations only:
\begin{equation}
S^{\mu \nu}P_{\nu}=0\,.
\end{equation}
Thus, using the above equations, we find that the non-vanishing components of the momentum are given by
\begin{equation}
\frac{P^t}{m}=\frac{e r^3-\frac{j s r^2}{2b}\left( a' b^2+2 a b b' \right)}{ a b^2 \Sigma }\,,
\end{equation}
\begin{equation}
\frac{P^{\phi}}{m}=\frac{jr-\frac{e s r}{b}}{ \Sigma}\,,
\end{equation}
and
\begin{equation}
\frac{P^r}{m}=\pm \sqrt{a^2b^2 \left(\frac{P^t}{m} \right)^2-ar^2 \left(\frac{P^{\phi}}{m}\right)^2-a}\,,
\end{equation}
where $e=E/m$ is the specific energy, $j=J/m$ is the total angular momentum per unit mass and $s=\pm S/m$ is the spin per unit mass. While, a positive value of the spin means that the spin is parallel to the total angular momentum, a negative value means that the spin is antiparallel to the total angular momentum. Also, $\Sigma$ is given by

\begin{eqnarray}
\notag \Sigma &=& r^3-\frac{s^2 r^2}{2}\left( a'+2 a \frac{ b' }{b} \right) \\
 &=& r^3-\frac{s^2}{2} \left( r \Delta'-\left( 1-2r\frac{b'}{b} \right)\Delta \right)\,,
\end{eqnarray}
where we have defined $\Delta = r a(r)$. Thus, by considering that the center of mass energy is given by
\begin{equation}
E^2_{CM}=-g_{\mu \nu} (P_1^{\mu} + P_2^{\mu})(P_1^{\nu}+P_2^{\nu})=m_1^2+m_2^2-2g_{\mu \nu}P_1^{\mu}P_2^{\nu}\, ,
\end{equation}
then the collisional energy for two spinning particles with the same mass $m=m_1=m_2$ in the background of the DST black hole yields

\begin{eqnarray}
\notag E^2_{CM} &=& \frac{2m^2}{\Delta \Sigma_1 \Sigma_2} \Bigg( \frac{r}{b^2} \left( e_1 r^3-\frac{j_1 s_1}{2} r^2 (a'b+2ab') \right) \left( e_2 r^3-\frac{j_2 s_2}{2} r^2 (a'b+2ab') \right)+ \\
\notag & & \Delta \left( \Sigma_1 \Sigma_2 -r^4 (j_1-e_1 s_1/b) (j_2-e_2 s_2/b) \right) - \\
\notag &&  \frac{1}{b^2} \sqrt{r \left( e_1 r^3 -\frac{j_1s_1}{2} r^2 (a'b+2ab') \right)^2 -\Delta b^2 \left( \Sigma_1^2+r^4 (j_1- e_1s_1/b)^2 \right)} \times   \\
&& \sqrt{r \left( e_2 r^3 -\frac{j_2s_2}{2} r^2 (a'b+2ab') \right)^2 -\Delta b^2 \left( \Sigma_2^2+r^4 (j_2- e_2s_2/b)^2 \right)}
\Bigg)\,. 
\end{eqnarray}
Possibles divergences can arise when the denominator of the above equation is zero, i.e., at the horizon radius $\Delta = 0$ and at a spin-related radius $\Sigma_i = 0$. In the first case, it can be demonstrated that the CM energy is finite at the horizon. In fact, setting $s_i=0$ for simplicity in the above expression with $i=1,2$, the $E^2_{CM}$ yields the finite value
\begin{equation}
\lim_{r \rightarrow r_H} \frac{E^2_{CM}}{m^2}= \frac{(e_2j_1-e_1 j_2)^2+(e_1+e_2)^2 r_H^2}{e_1 e_2 r_H^2}\,.
\end{equation}
In the case $s_i \neq 0$ is more difficult to find an analytic expression like the  above expression; However, it can be shown that $E^2_{CM}$ also is finite when $\Delta \rightarrow 0$. For instance, in Fig. (\ref{CM}) we can see that the CM energy does not diverges at the horizon for some particular values of the parameters. On the other hand, when $\Sigma_i \rightarrow 0$ the $E^2_{CM}$ diverges, indicating an infinite CM energy at the spin-related radius, this behavior is shown in the same Fig. (\ref{CM}) for some values of the parameters, which show that the divergence of the CM energy occurs outside the event horizon. However, it must be analyzed if the spinning particles can reach the divergence radius. In \cite{Armaza:2015eha} was showed that for the Schwarzschild black hole some particles with retrograde orbits in principle can reach the divergence radius before reaching the horizon.
Now, rewriting $P^r$ as
\begin{equation}
\left( \frac{P^r}{m} \right)^2= \frac{1-\frac{as^2}{r^2}}{b^2 \left( 1-\frac{s^2}{2r}(a'+2a \frac{b'}{b}) \right)^2}(e-V_+)(e-V_-)\,,
\end{equation}
where the effective potentials $V_{\pm}$ are defined as
\begin{equation}
V_{\pm}=\left(1-\frac{as^2}{r^2} \right)^{-1}\Bigg( (a'br^2+2ab'r^2-2abr) \frac{js}{2r^3} \pm \sqrt{a \left( 1+\frac{j^2}{r^2}-\frac{as^2}{r^2} \right) \left( 1-\frac{s^2}{2r} \left(a'+2a \frac{b'}{b} \right) \right)^2}\Bigg)\,.
\end{equation}
Notice that for the case $1-as^2/r^2>0$, $e$ must be bigger than $V_+$ or smaller than $V_-$ for a real value of $P^r$.
Therefore, from Eq. \eqref{motion} and Eq. \eqref{motion1} we obtain the following set of momentum equations
\begin{eqnarray}
\notag && \dot{P}^t+\left( \frac{a'}{2a}+\frac{b'}{b} \right)  P^t \dot{r} + \left( \frac{a'}{2a} +\frac{b'}{b} \right) P^r \dot{t} = \left( \frac{3 a' b'}{2ab}+ \frac{a''}{2 a}+\frac{b''}{b} \right) S^{tr} \dot{r} +\left( a'+2 a\frac{b'}{b} \right) \frac{r}{2} S^{t \phi} \dot{\phi} \\
\notag && \dot{P}^r+\frac{1}{2}ab^2 \left( a'+2 a \frac{b'}{b} \right) P^t \dot{t} -\frac{a'}{2a}P^r \dot{r}-r aP^{\phi} \dot{\phi} = a^2 b^2 \left( \frac{3 a' b'}{2ab}+ \frac{a''}{2 a}+\frac{b''}{b} \right) S^{tr} \dot{t} +\frac{1}{2} r a' S^{r \phi} \dot{\phi} \\
&& \dot{P}^{\phi}+\frac{1}{r}P^r \dot{\phi} +\frac{1}{r} P^{\phi} \dot{r}=-\frac{a'}{2 r a}S^{r \phi} \dot{r} +\frac{a b^2}{2r} \left(a'+2a \frac{b'}{b}\right) S^{t \phi} \dot{t}
\end{eqnarray}
and the spin equations are given by
\begin{eqnarray}
\notag  && \dot{S}^{tr}+\frac{b'}{b} S^{tr} \dot{r} -raS^{t \phi} \dot{\phi}=P^t \dot{r}-P^r \dot{t}  \\
\notag  && \dot{S}^{r \phi} +\frac{1}{2}ab^2 \left( a'+2a\frac{b'}{b} \right) S^{t \phi} \dot{t}+\left(-\frac{a'}{2a} +\frac{1}{r} \right) S^{r \phi} \dot{r} = P^r \dot{\phi}- P^{\phi} \dot{r} \\
 && \dot{S}^{t \phi}+ \frac{1}{r} S^{tr} \dot{\phi}+ \left( \frac{a'}{2a}+\frac{b'}{b} +\frac{1}{r} \right) S^{t \phi} \dot{r} +\left( \frac{a'}{2a} +\frac{b'}{b} \right) S^{r \phi} \dot{t}= P^t \dot{\phi} -P^{\phi} \dot{t}\,.
\end{eqnarray}
These equations imply

\begin{equation}
\dot{r}= \frac{1-\frac{a's^2}{2r}-\frac{as^2 b'}{rb}}{1-\frac{a's^2}{2r}} \frac{P^r}{P^t} \dot{t}
\end{equation}
\begin{equation}
\dot{\phi}=-\frac{-2+s^2 a'' +3a' \frac{b'}{b} s^2+2 a \frac{b''}{b}s^2}{2 \left( 1-\frac{a's^2}{2r} \right)}\frac{P^{\phi}}{P^t} \dot{t}\,.
\end{equation}
Using the above expressions we can evaluate the velocity square
\begin{equation}
\frac{u_{\mu}u^{\mu}}{\left( u^t \right)^2}=-a(r)b(r)^2+\frac{1}{a(r)} \left( \frac{\dot{r}}{\dot{t}} \right)^2+r^2\left( \frac{\dot{\phi}}{\dot{t}} \right)^2\,.
\end{equation}
In Fig. \ref{uu} we plot the behavior of $u_{\mu}u^{\mu}/ \left( u^t \right)^2$ and $\Sigma$ as a function of $r$ and $s$ for some small values of $\sigma$. The green surface corresponds to the behavior of $u_{\mu}u^{\mu}/ \left( u^t \right)^2$ and the red surface  corresponds to the behavior of $\Sigma$. The intersection between the green surface and the $z=0$ horizontal plane is the limit where the trajectories change from  timelike to spacelike character. The intersection between the red surface and the $z=0$ horizontal plane corresponds to the values for which the CM energy diverges. We can observe that in order to reach a divergence in the energy of the center mass  that the trajectory of the STOP has to pass from timelike to spacelike, which is similar to the collisions of spinning particles in the Schwarzschild background \cite{Armaza:2015eha}. We recover the case for the Schwarzschild black hole when $\sigma=0$, see the central panel of Fig. (\ref{uu}).

\newpage

\section{Conclusions}
\label{conclusion}

In this paper we studied the motion of particles in the background of a DST black hole. We analyzed the motion of particles in the equatorial plane and we recovered the classical result of GR for $\sigma=0$, and $r_S=2M$ \cite{chandra}, recall $\sigma=0$ corresponds to $\beta_n=0$ due to its definition.
A qualitative analysis of the effective potential for null geodesics shows that  behavior for radial photons is  similar to those in a  Schwarzschild's spacetime \cite{chandra}. The same occurs for the motion of photons with angular momentum, when $-1/2 < \sigma < 1/4$. Where, unstable circular orbits depend on the coupling parameter $\sigma$.  However, for $\sigma<-1/2$, all orbits  are bounded, which does not occur in the Schwarzschild's spacetime. The discrepancy between the theoretical value and
the observational value of the deflection light was studied with respect to small deviations of Schwarzschild's spacetime. We have found via the geodesics formalism that 
the value of the coupling constant $\beta_2$ to match the theoretical result with the current observational constrains is $-8.97241\times 10^{-6} \sqrt{3}\leq \beta_2 \leq 3.41708\times10^{-6} \sqrt{3}$. \\

Through the study for radial motion of massive particles we obtain new geodesics: for $\sigma<0$ all  orbits are bounded and for $0<\sigma<1/4$ appears an unstable equilibrium point ($r_u$), and two critical trajectories  approaching  to this point asymptotically with the same energy $E_u$. For particles with $E>E_u$, the trajectories are unbounded. For $E<E_u$, we showed that they allow a frontal scattering, that is characterized for incoming particles  from infinity to a radial distance of closest approach, and come back to the infinity. With respect to the motion of particles with  angular momentum and $\sigma<0$, all trajectories are bounded due to potential at infinity, 
as for  Schwarzschild AdS black hole \cite{Cruz:2004ts} and the classical GR orbits are also allowed. Interestingly, for $0<\sigma<1/4$, the spacetime has two unstable circular orbits and one stable circular orbit, and as the potential vanishes at infinity,
not all the orbits are bounded. Also, there are planetary orbits, which  allowed us to study the perihelion precession for Mercury and the discrepancy between
 the theoretical and observational value,  for small deviations of Schwarzschild's spacetime. In this regards, we found that it is possible attribute the discrepancy to a DST theory with a coupling constant $\beta_2 = 1.244 * 10 ^{-9}\sqrt{3}$. Due to the value of the coupling constant found for the perihelion precession is contained in the range obtained for the bending of light, these observables can be predicted by a unique DST theory according to observational data. \\

In consideration of  the collision of spinning particles near the horizon and the possibility that the DST black hole acts as a particle accelerator, we showed  that to reach a divergence in the center mass energy, 
the trajectory of the STOP has to pass from timelike to spacelike. Thus, for small  deviations of General  Relativity, the behavior is similar to the one observed for collisions of spinning particles in the Schwarzschild background \cite{Armaza:2015eha}.

\acknowledgments
We would like to thank Thomas M\"adler for his comments and suggestions. This work is supported by Comisi\'on Nacional
de Ciencias y Tecnolog\'ia of Chile through FONDECYT Grant N$^{\textup{o}}$ 3170035 (A. \"{O}.), N$^{\textup{o}}$ 1170279 (J. S.) and by the Direcci\'{o}n de Investigaci\'{o}n y Desarrollo de la Universidad de La Serena (Y.V.). P. A. G. acknowledges the hospitality of the Universidad de La Serena where part of this work was undertaken. A. \"{O}. is  grateful to Institute for Advanced Study, Princeton for hospitality.

\end{document}